\documentclass[11pt]{article}

\usepackage{amsmath}
\usepackage{amssymb}
\usepackage{amsthm}
\usepackage{latexsym}
\usepackage{color}
\usepackage{graphicx}
\usepackage{appendix}
\usepackage{color} 
\usepackage{enumerate}
\usepackage[T1]{fontenc}
\usepackage[english]{babel}
\usepackage[table]{xcolor}

\DeclareSymbolFont{calletters}{OMS}{cmsy}{m}{n}
\DeclareSymbolFontAlphabet{\mathcal}{calletters}

\def\be{\begin{eqnarray}}
\def\ee{\end{eqnarray}}

\def\b*{\begin{eqnarray*}}
\def\e*{\end{eqnarray*}}

\newtheorem{Theorem}{Theorem}[part]
\newtheorem{Definition}{Definition}[part]
\newtheorem{Proposition}{Proposition}[part]

\newtheorem{Lemma}{Lemma}[part]

\newtheorem{Remark}{Remark}[part]
\newtheorem{Example}{Example}[part]

\newtheorem{Condition}{Condition}[part]

\makeatletter \@addtoreset{equation}{section}

\@addtoreset{Definition}{section}

\@addtoreset{Theorem}{section}

\@addtoreset{Proposition}{section}

\@addtoreset{Property}{section}

\@addtoreset{Assumption}{section}

\@addtoreset{Corollary}{section}

\@addtoreset{Lemma}{section}

\@addtoreset{Remark}{section}

\@addtoreset{Example}{section}

\@addtoreset{Condition}{section}

\newcommand{\abs}[1]{\left|#1\right|}     

\def \E{\mathbb{E}}
\def \F{\mathbb{F}}
\def \H{\mathbb{H}}
\def \L{\mathbb{L}}

\def \P{\mathbb{P}}

\def \R{\mathbb{R}}

\def \X{\mathbb{X}}
\def \G{\mathbb{G}}

\def\esup{{\rm ess \, sup}}

\def\no{\noindent}

\def\={\;=\;}
\def\.{\;.}

\def\eps{\varepsilon}

\def\reff#1{{\rm(\ref{#1})}}

\def\1{{\bf 1}}

\def\eps{\epsilon}

\def\b*{\begin{eqnarray*}}
\def\e*{\end{eqnarray*}}

 \def\normeL2#1{\left\|{#1}\right\|_{L^2}}
 
 \title{On the Robust superhedging of measurable claims}

 \author{Dylan {\sc Possama\"{i}} \footnote{CEREMADE, Universit\'e Paris Dauphine, possamai@ceremade.dauphine.fr.}
 \and Guillaume {\sc Royer}\footnote{CMAP, Ecole Polytechnique Paris, guillaume.royer@polytechnique.edu.}
      \and Nizar {\sc Touzi}\footnote{CMAP, Ecole Polytechnique Paris, nizar.touzi@polytechnique.edu.
      Research supported by the Chair {\it Financial Risks} of the {\it Risk Foundation} sponsored by Soci\'et\'e
             G\'en\'erale, and
             the Chair {\it Finance and Sustainable Development} sponsored by EDF and Calyon. }}
             \date{\today}

\begin{document}

 \maketitle

 \begin{abstract}

The problem of robust hedging requires to solve the problem of superhedging under a nondominated family of singular measures. Recent progress was achieved by \cite{nvh,nn}. We show that the dual formulation of this problem is valid in a context suitable for martingale optimal transportation or, more generally, for optimal transportation under controlled stochastic dynamics.

\vspace{10mm}

\vspace{5mm}

\noindent{\bf AMS 2000 subject classifications:} 

\end{abstract}

\section{Introduction}
An important attention is focused on the problem of robust superhedging in the recent literature. Motivated by the original works of Avellaneda \cite{Avellaneda} and Lyons \cite{Lyons}, the first general formulation of this problem was introduced by Denis and Martini \cite{DM} by considering the hedging problem under a nondominated family of probability measures on the canonical space of continuous trajectories. Since the hedging problem involves stochastic integration, \cite{DM} used the capacity theory to develop the corresponding quasi-sure stochastic analysis tools, i.e. stochastic analysis results holding simultaneously under the considered family non-dominated measures. 

The next progress was achieved by Soner, Touzi and Zhang \cite{stz} who introduced a restriction of the set of non-dominated measures so as to guarantee that the predictable representation property holds true under each measure. However, \cite{stz} placed strong regularity conditions on the random variables of interest in order to guarantee the measurability of the value function of some dynamic version of a stochastic control problem, and to derive the corresponding dynamic programming principle. 

By using the notion of measurable analyticity, van Handel and Nutz \cite{nvh} and Neufeld and Nutz \cite{nn} extended the previous results to general measurable claims by introducing some conditions that the non-dominated family of singular measures must satisfy.

The main objective of this paper is to extend the approach of Neufeld and Nutz \cite{nn} so as to introduce some specific additional constraints on the family of probability measures, and to weaken the integrability condition on the random variables of interest. Such an extension is crucially needed in the recent problem of martingale transportation problem \cite{ght,host}, where the superhedging problem allows for the static trading of any Vanilla payoff in addition to the dynamic trading of the underlying risky asset. Assuming that the financial market, with this enlarged possibilities of trading, satisfies the no-arbitrage condition leads essentially to the restriction of the family of probability measures to those under which the canonical process is a uniformly integrable martingale. The main problem is that this restriction violates the conditions of \cite{nn} on one hand, and that the integrability conditions in \cite{nn} are not convenient for the stochastic control approach of \cite{ght,host}. 

The paper is organized as follows. Section \ref{sect:preliminary} introduces the main probabilistic framework. The robust superhedging problem is formulated in Section \ref{sect:mainresult}, where we also report our main result, together with the comparison to \cite{nn}. Section \ref{sect:duality} contains the proof of the duality result. Finally, some extensions are reported in Section \ref{sect:extensions}. 

\section{Preliminaries}
\label{sect:preliminary}
\subsection{Probabilistic framework}
Let $ \Omega \mathrel{\mathop:}= \lbrace \omega \in C\left( \left[ 0,T \right], \mathbb{R}^d \right) : \omega_0 = 0 \rbrace $ be the canonical space equipped with the uniform norm $ ||\omega||_{\infty}^T \mathrel{\mathop:}= \sup_{0\le t \le T} |\omega_t| $. $\mathcal F$ will always be a fixed $\sigma$-field on $\Omega$ which contains all our filtrations. We then denote $ B $ the canonical process, $ \mathbb{P}_0 $ the Wiener measure, $ \mathbb{F} \mathrel{\mathop:}= \lbrace \mathcal{F}_t \rbrace_{0\le t \le T} $ the filtration generated by $ B$ and $ \mathbb{F}^+ \mathrel{\mathop:}= \lbrace \mathcal{F}^+_t , 0 \le t \le T \rbrace $, the right limit of  $ \mathbb{F} $ where $ \mathcal{F}_t^+ \mathrel{\mathop:}= \cap_{ s> t} \mathcal{F} $. We will denote by $\mathbf{M}(\Omega)$ the set of all probability measures on $\Omega$. We also recall the so-called universal filtration $\F^*:=\lbrace \mathcal{F}^*_t \rbrace_{0\le t \le T} $ defined as follows
$$
\mathcal F^*_t
:=
\underset{\mathbb P\in\mathbf{M}(\Omega)}
{\bigcap}\mathcal F_t^{\mathbb P},
$$ 
where $\mathcal F_t^{\mathbb P}$ is the usual completion under $\mathbb P$. 

\vspace{0.3em}
For any subset $E$ of a finite dimensional space and any filtration $\mathbb X$ on $(\Omega,\mathcal F)$, we denote by $\H^0(E,\mathbb X)$ the set of all $\X$-progressively measurable processes with values in $E$. Moreover for all $p>0$ and for all $\P\in\mathbf M(\Omega)$, we denote by $\H^p(\P,E,\X)$ the subset of $\H^0(E,\X)$ whose elements $H$ satisfy $\E^\P\left[\int_0^T\abs{H_t}^pdt\right]<+\infty.$ The localized versions of these spaces are denoted by $\H^p_{\rm loc}(\P,E,\X)$. 

\vspace{0.3em}
For any subset ${\cal P}\subset\mathbf{M}(\Omega)$, a ${\cal P}-$polar set is a $\P-$negligible set for all $\P\in{\cal P}$, and we say that a property holds ${\cal P}-$quasi-surely if it holds outside of a ${\cal P}-$polar set. Finally, we introduce the following filtration $\mathbb G^{\mathcal P}:=\lbrace \mathcal{G}^\mathcal P_t \rbrace_{0\le t \le T} $ which will be useful in the sequel
$$\mathcal G^{\mathcal P}_t:=\mathcal F_{t^+}^*\vee \mathcal N^\mathcal P,\ t< T\text{ and }\mathcal G^{\mathcal P}_T:=\mathcal F_{T}^*\vee \mathcal N^\mathcal P,$$
where $\mathcal N^\mathcal P$ is the collection of $\mathcal P$-polar sets.

\vspace{0.3em}

For all  $\alpha\in\H^1_{\rm loc}(\P_0,\mathbb{S}^{>0}_d,\F) $, where $ \mathbb{S}^{>0}_d$ is the set of positive definite matrices of size $d\times d$, we define the probability measure on $ (\Omega, \mathcal{F} )$
$$ \mathbb{P}^{\alpha}:= \mathbb{P}_0 \circ (X^{\alpha}_.)^{-1} \text{ where } X^{\alpha}_t := \int_0^t \alpha_s^{1/2} dB_s, \ t \in [0,T], \ \mathbb{P}_0-a.s.$$
We denote by $ \mathcal{P}_S $ the collection of all such probability measures on $ (\Omega,\mathcal{F})$. We recall from Karandikar \cite{k} that the quadratic variation process $ \left \langle B \right \rangle $ is universally defined under any $ \mathbb{P} \in \mathcal{P}_S $, and takes values in the set of all nondecreasing continuous functions from $ \mathbb{R}_+ $ to $ \mathbb{S}_d^{>0}$. We will denote its pathwise density with respect to the Lebesgue measure by $\widehat a$. Finally we recall from \cite{stz-quasi} that every $ \mathbb{P} \in \mathcal{P}_S $ satisfies the Blumenthal zero-one law and the martingale representation property.

\vspace{0.3em}

Our focus in this paper will be on the following subset of $\mathcal{P}_S$.

\begin{Definition}
$\mathcal{P}_m$ is the sub-class of $ \mathcal{P}_S$ consisting of all $ \mathbb{P} \in \mathcal{P}_S $ such that the canonical process $B$ is a $\mathbb{P}-$uniformly integrable martingale.
\end{Definition}

\subsection{Regular conditional probability distributions}

In this section, we recall the notion of regular conditional probability distribution (r.c.p.d.), as introduced by Stroock and Varadhan \cite{sv}. Let $\mathbb{P}\in\mathbf{M}(\Omega)$ and consider some $\mathbb{F}$-stopping time $\tau$. Then, for every $\omega \in\Omega$, there exists an r.c.p.d. $ \mathbb{P}^{\omega}_{\tau} $ satisfying:
\\
(i) $\mathbb{P}^{\omega}_{\tau} $ is a probability measure on $ \mathcal{F}_T $.
\\	
(ii) For each $ E \in \mathcal{F}_T $, the mapping $ \omega \rightarrow \mathbb{P}^{\omega}_{\tau}(E) $ is $ \mathcal{F}_\tau $-measurable.
\\
(iii) $ \mathbb{P}^{\omega}_{\tau} $ is a version of the conditional probability measure of $ \mathbb{P} $ on $ \mathcal{F}_{\tau}$, i.e., for every integrable $ \mathcal{F}_T $-measurable r.v. $ \xi $ we have
$ \mathbb{E}^{\mathbb{P} } [ \xi | \mathcal{F}_{\tau}](\omega)=\mathbb{E}^{ \mathbb{P}^{\omega}_{\tau}} [\xi],$ $\P-$a.s.
\\
(iv) $ \mathbb{P}^{\omega}_{\tau} (\Omega_{\tau}^{\omega})=1$, where $\Omega_{\tau}^{\omega}\mathrel{\mathop:}=\lbrace \omega' \in \Omega : \ \omega'(s)=\omega(s), \ 0\le s \le \tau(\omega) \rbrace.$

\vspace{0.3em}
We next introduce the shifted canonical space and the corresponding notations.

\vspace{0.3em}

$\bullet$ For $ 0 \le t \le T $, denote by $ \Omega^t \mathrel{\mathop:}=\lbrace \omega \in C([t,T],\mathbb{R}) \ : \ w(t)=0 \rbrace  $ the shifted canonical space, $ B^t $ the shifted canonical process on $ \Omega^t $, $ \mathbb{P}_0^t $ the shifted wiener measure, $ \mathbb{F}^t $ the shifted filtration generated by $ B^t $.

\vspace{0.3em}
$\bullet$ For $ 0 \le s \le t \le T $, $ \omega \in \Omega^s $, define the shifted path $ \omega^t \in \Omega^t $, $ \omega_r^t \mathrel{\mathop:}= \omega_r -\omega_t \ \text{for all} \  \ r \in [t,T]. $

\vspace{0.3em}
$\bullet$ For  $ 0 \le s \le t \le T $, $ \omega \in \Omega^s $, define the concatenation path $ \omega \otimes_t \widetilde{\omega} \in \Omega^s $ by:
$$ ( \omega \otimes_t \widetilde{\omega} ) (r) \mathrel{\mathop:}= \omega_r {\bf{1}}_{ [s,t)}(r) + (\omega_t + \widetilde{\omega}_r){\bf{1}}_{[t,1]}(r) \ \ \text{for all} \  \ r \in [s,T].$$

\vspace{0.3em}
$\bullet$ For $ 0 \le s \le t \le T $, for any $ \mathcal{F}_T^s$-measurable random variable $ \xi $ on $ \Omega^s $, and for each $ \omega \in \Omega^s $, define the shifted $ \mathcal{F}_T^t $-measurable random variable $ \xi^{t, \omega} $ on $ \Omega^t $ by:
$$ \xi^{t,\omega}(\widetilde{\omega} ) \mathrel{\mathop:}= \xi (\omega \otimes_t \widetilde{\omega} ) \ \ \text{for all} \ \  \widetilde{\omega} \in \Omega^t. $$

\vspace{0.3em}
$\bullet$ The r.c.p.d. $ \mathbb{P}_{\tau}^{\omega} $ induces naturally a probability measure $ \mathbb{P}^{\tau ,\omega} $ on $ \mathcal{F}^{\tau(\omega)}_T $ such that the $ \mathbb{P}^{\tau , \omega} $-distribution of $ B^{\tau(\omega)}$  is equal to the $ \mathbb{P}_{\tau}^{\omega} $-distribution of $ \lbrace B_t- B_{\tau(\omega)} , \ t \in [\tau(\omega) , T ) \rbrace $. It is then clear that for every integrable and $ \mathcal{F}_T$-measurable random variable $ \xi $,
$$ \E^{\mathbb{P}_{\tau}^{\omega}}[\xi] = \E^{\mathbb{P}^{\tau ,\omega}}[\xi^{\tau , \omega} ].$$ 
For the sake of simplicity, we shall also call $ \mathbb{P}^{\tau ,\omega} $ the r.c.p.d. of  $ \mathbb{P} $.

\vspace{0.3em}
$\bullet$ Finally, we introduce for all $ (s,\omega)\in [0,T] \times \Omega$:
\begin{align*}
\mathcal{P}_S(s,\omega) \mathrel{\mathop:}= 
\left\{ \mathbb{P}_0^s \circ \Big(\int_s^{\cdot}\alpha^{1/2}_u dB_u^s\Big)^{-1}, \ \text{with} \ \int_s^T|\alpha_u|du < + \infty , \ \mathbb{P}_0^s-\text{a.s.} \right\}\\
\mathcal P_m(s,\omega):=\left\{\mathbb P\in\mathcal P_s(s,\omega)\text{ s.t. }B^s\text{ is a uniformly integrable martingale}\right\}.
\end{align*}
It is clear that the families $ (\mathcal{P}_S(s,\omega))_{(s,\omega)\in [0,T]\times \Omega}$ and $(\mathcal{P}_m(s,\omega))_{(s,\omega)\in [0,T]\times \Omega}$ are adapted in the sense that $\mathcal{P}_S(s,\omega)=\mathcal{P}_S(s,\widetilde{\omega})$ and $\mathcal{P}_m(s,\omega)=\mathcal{P}_m(s,\widetilde{\omega})$, whenever $\omega|_{[0,s]}=\widetilde{\omega}|_{[0,s]}$.

\section{Superreplication and duality}
\label{sect:mainresult}

\subsection{Problem formulation and main results}

Throughout this paper, we consider some scalar $\mathcal{G}_T$-measurable random variable $\xi$. For any $(s,\omega)\in[0,T]\times\Omega$, we naturally restrict the subset $\mathcal P_m$ and $\mathcal P_m(s,\omega)$ to:
 \begin{align*}
 {\cal P}_m^\xi
 &:=
 \big\{\P\in\mathcal P_m:~\E^\P[\xi^-]<+\infty\big\}\\
  {\cal P}_m^\xi(s,\omega)
 &:=
 \big\{\P\in\mathcal P_m(s,\omega):~\E^\P[(\xi^{s,\omega})^-]<+\infty\big\}.
 \end{align*}
 Notice that such a restriction can be interpreted as suppressing measures which induce arbitrage opportunities in our market.
 
 \vspace{0.3em}
Our main interest is on the problem of superreplication under model uncertainty and the corresponding dual formulation. Given some initial capital $X_0$, the wealth process is:
 $$ 
 X_t^H \mathrel{\mathop:}= X_0 + \int_0^t H_s dB_s, \ t \in [0,T],
 $$
where $H\in{\cal H}^\xi$, the set of admissible trading strategies defined by:
 \begin{equation*}
 {\cal H}^\xi
 :=
 \left\{H\in\H^0(\R^d,\G^{\mathcal P_m})\cap\H^2_{\text{loc}}(\P,\R^d,\G^{\mathcal P_m}), ~X^H~\mbox{is a}~\P-\mbox{supermartingale, $\forall$}~\P\in{\cal P}_m^\xi
 \right\}.
 \end{equation*}
The main result of this paper is the following.

\begin{Theorem}\label{th.main}
Let $\xi$ be an upper semi-analytic r.v. with $\sup_{\P\in\mathcal{P}_m}\E^{\P}[\xi^+]<+\infty $. Then
 \b*
 V(\xi) 
 &\mathrel{\mathop:}=& 
 \inf \left\{X_0:~X^H_T \ge \xi, \ \ \mathcal P_{m}^\xi-
                 \mbox{q.s. for some}~H\in{\cal H}^\xi 
      \right\} 
 \;=\; 
 \sup_{\P \in \mathcal{P}_m} \E^{\P}[\xi].
 \e*
Moreover, existence holds for the primal problem, i.e. $V(\xi)+\int_0^T H_sdB_s\ge\xi$, ${\cal P}_m^\xi-$q.s. for some $H\in{\cal H}^\xi$.
\end{Theorem}

\begin{Remark}
Suppose that the random variable $\xi^-$ is $\P-$integrable for all $\P\in{\cal P}_m$. Then, ${\cal P}_m^\xi={\cal P}_m$, and the corresponding set of admissible strategies ${\cal H}^\xi=:{\cal H}$ is independent of $\xi$. Under the condition $\sup_{\P\in{\cal P}_\infty}\E^\P[\xi^+]<\infty$, it follows from the previous theorem that:
 \b*
 \inf \left\{X_0:~X^H_T \ge \xi, \ \ \mathcal P_{m}-
                 \mbox{q.s. for some}~H\in{\cal H}
      \right\}
 \;=\; 
 \sup_{\P \in \mathcal{P}_m} \E^{\P}[\xi].
 \e*
\end{Remark}

\begin{Remark}
When it comes to which filtration the trading strategies are admissible, we can actually do a little bit better than $\G^{\mathcal P_m}$, and consider the universal filtration $\F^*$ completed by the $\mathcal P_m$-polar sets, instead of its right limit. Indeed, let $X$ be a process adapted to $\G^{\mathcal P_m}$, then following the arguments in Lemma $2.4$ of \cite{stz-quasi}, we define $\widetilde X$ by
$$\widetilde X_t:=\underset{\eps\downarrow 0}{\limsup}\ \frac1\eps\int_{t-\eps}^tX_sds.$$
Then, $\widetilde X$ coincides $dt\times\mathbb P-$a.e. with $X$, for any $\P\in\mathcal P_S$ and is adapted to $\F^*$ completed by the $\mathcal P_m$-polar sets. For simplicity, we however refrain from considering this extension.\footnote{We would like to thank Marcel Nutz for pointing this out.}
\end{Remark}

The problem of superhedging under model uncertainty was first considered by Denis and Martini \cite{dm} using the theory of capacities and the quasi-sure analysis. The set of probability measures considered in \cite{dm} is larger than $\mathcal P_S$, and whether existence of an optimal hedging strategy strategy holds or not in the framework of \cite{dm} is still an open problem. Later, Soner, Touzi and Zhang \cite{stz-quasi} considered the same problem but with a strict subset of $\mathcal P_S$ satisfying a separability condition, which allowed them to recover the existence of an optimal strategy. The same approach is adapted in Galichon, Henry-Labord\`ere and Touzi to obtain the duality result of Theorem \ref{th.main} for uniformly continuous $\xi$. Recently, Neufeld and Nutz \cite{nn} introduced a new approach which avoids the strong regularity condition on $\xi$. In the next subsection, we briefly outline their approach and explain why it fails to cover our framework.  

\subsection{The analytic measurability approach}

We now introduce the general framework of \cite{nvh} and \cite{nn}. Let  $\mathcal P$ be a non-empty subset of $\mathcal P_S$, with corresponding "shifted" sets $\mathcal P(s,\omega)$, satisfying:

\begin{Condition} \label{cond.nutz}
Let $s\in \mathbb{R}_+$, $ \tau$ a stopping time such that $ \tau \geq s$, $\overline{\omega}\in \Omega$, and $ \mathbb{P}\in \mathcal{P}(s,\overline{\omega})$. Set $ \theta \mathrel{\mathop:}=\tau^{s,\overline{\omega}}-s $.
\\
{\rm{(i)}} The graph $ \lbrace (\mathbb{P}',\omega): \omega \in \Omega, \ \mathbb{P}' \in \mathcal{P}(t,\omega) \rbrace \subseteq \mathbf{M}(\Omega) \times \Omega $ is analytic.
\\
{\rm{(ii)}}  We have $ \mathbb{P}^{\theta,\omega} \in \mathcal{P}(\tau,\overline{\omega} \otimes_s \omega ) $ for $ \mathbb{P}$-a.e. $ \omega \in \Omega $.
\\
{\rm{(iii)}} If $ \nu : \Omega \rightarrow \mathcal{B}(\Omega)$ is an $ \mathcal{F}_{\theta}$-measurable kernel and $ \nu(\omega) \in \mathcal{P}(\tau,\overline{\omega}\otimes_s \omega)$ for $ \mathbb{P}$-a.e. $ \omega \in \Omega $, then the following measure $\overline{\mathbb P}\in \mathcal{P}(s,\overline{\omega})$:
$$\overline{\mathbb P}(A):= \iint ({\bf{1}}_A)^{\theta,\omega} ( \omega') \nu(d \omega' ; \omega ) \mathbb{P}(dw), \ \ A \in \mathcal{F}.$$
\end{Condition}

\begin{Theorem}[Theorem 2.3 in \cite{nn}]
Suppose $\{\mathcal P(s,\omega)\}_{(s,\omega)}$ satisfies Condition \ref{cond.nutz}. Then, for any upper semi-analytic map $\xi$ with $\sup_{\mathbb P\in\mathcal P}\mathbb E^\mathbb P[\abs{\xi}]<+\infty$, we have:
 \b*
 \inf\left\{X_0:~X^H_T \ge \xi \ \ {\mathcal P}-
                \mbox{q.s. for some}~ H \in \mathcal{H}
     \right\} 
 &=& 
 \underset{\P \in \mathcal{P}}{\sup}\  \E^{\P}[\xi].
 \e*
\end{Theorem}

For the purpose of the application of this result to the problem of martingale optimal transportation, see \cite{ght,host}, the last result presents two inconveniences:
\\
- The integrability condition of the previous theorem from \cite{nn} turns out to be too strong. The weaker integrability conditions in our Theorem \ref{th.main} is a crucial for the analysis conducted in \cite{ght,host}. \\
- The set of probability measures of interest is the smaller subset ${\mathcal P}_m$. We shall verify below that ${\mathcal P}_m$ satisfies Conditions \ref{cond.nutz}(i) and (ii), but fails to satisfy (iii). Therefore, we need to extend the results of \cite{nn} in order to address the case of ${\mathcal P}_m$.

\begin{Example}{\rm[${\mathcal P}_m$ does not satisfy Condition \ref{cond.nutz} (iii)]} \label{example}
For simplicity, let $d=1$. Let $ s \in (0,T) $, $t \geq s$, $ \overline{\omega} \in\Omega$ and $ \mathbb{P}=\P_0^s \in \mathcal{P}_{B}(s,\overline{\omega})$. Now consider $ \omega \in \Omega^s $. The family $ ( \mathbb{P}_i)_{i\in \mathbb{N} } $ is defined by
$$ \forall i \in \mathbb{N} , \ \mathbb{P}_i=\mathbb{P}_0^t \circ \left( \int_0^{\cdot} \sigma_i^{1/2} dB_u^t \right)^{-1}.$$
where $ (\sigma_i)_{i\in \mathbb{N}}$ is a sequence of positive numbers which will be chosen later. We consider the following partition $ (E_i)_{i\in \mathbb{N}}$ of $ \mathcal{F}_t$
$$ \forall i \in \mathbb{N}, \ E_i\mathrel{\mathop:}= \lbrace \omega \ \text{s.t.} \ \omega_t \in (-i-1,-i] \cup [i,i+1)\rbrace .$$
We then introduce the $ \mathcal{F}_t$-measurable kernel $ \nu(\omega)(A)\mathrel{\mathop:}=\sum_{i=0}^{+\infty} {\bf{1}}_{E_i}(\omega) \mathbb{P}_i(A), $ and we define $ \overline{\P} $ as in Condition \ref{cond.nutz}(iii) from $\mathbb P$ and $\nu$. We now show that $ \E^{\overline{\P}} [|B_T|]=+\infty$ for some convenient choice of the sequence $ ( \sigma^i)$. In particular this shows that $ \overline{\P}\notin \mathcal{P}_m $.
\begin{align*}
\E^{\overline{\mathbb{P}}}[|B_T|] 
= 
\E^{\mathbb{P}}\left[\E^{\overline{\mathbb{P}}}[|B_T| | \mathcal{F}_t]\right] &=  \E^{\mathbb{P}}\left[\E^{\overline{\mathbb{P}}^{t,\overline{\omega}\otimes_s \omega}}\left[|B_T|\right]\right] \\
&=
\E^{\mathbb{P}}\left[\sum_{i=0}^{+\infty} {\bf{1}}_{E_i}(\omega) \E^{\mathbb{P}_i}\left[|B_T^{t,\overline{\omega}\otimes_s \omega}|\right]\right] \\
&=
\E^{\mathbb{P}}\left[\sum_{i=0}^{+\infty} {\bf{1}}_{E_i}(\omega)\int_{-\infty}^{+\infty}|\sigma_i u +B_t(\omega)| \frac{e^{-u^2/2}}{\sqrt{2 \pi}}du\right] 
=
\sum_{i=0}^{+\infty} f_i(\sigma_i),
\end{align*}
where, for all $i$,
$$ f_i(\sigma)
:= \frac{\sigma}{2\pi} \left(  \int_{-i-1}^{-i}\int_{-\infty}^{+\infty}\abs{u +\frac{ty}{\sigma}} e^{-\frac{u^2+y^2}{2}}du dy  + \int_{i}^{i+1}\int_{-\infty}^{+\infty}\abs{u +\frac{ty}{\sigma}} e^{-\frac{u^2+y^2}{2}}du dy   \right). 
$$
Notice that $f_i(\sigma)\longrightarrow\infty $, as $\sigma \rightarrow \infty$. Then there exists $ \sigma_i >0 $ such that $ f_i(\sigma_i) \geq 1 $. Hence, $ \E^{\overline{\mathbb{P}}}[|B_T|]=+\infty$, where $\overline{\mathbb{P}}$ is defined using this family of coefficients.
\end{Example}

\begin{Proposition}\label{prop.i-ii}
$ \mathcal{P}_m $ and $\mathcal P_m^\xi$ verify Condition \ref{cond.nutz} (i) and (ii).
\end{Proposition}

\vspace{0.3em}
\textbf{Proof} We only provide the proof for $\mathcal P_m$, the result for $\mathcal P_m^\xi$ follows by direct adaptation. We first verify Condition \ref{cond.nutz} (ii). Let $\P \in \mathcal{P}_m$, and consider an arbitrary $\mathbb F$-stopping time $\tau$, and $\mathbb F^\tau$-stopping time $\sigma$. By Lemma A.1 in \cite{kpz}, there exists some $\mathbb F^\tau$-stopping time $\tilde\sigma$ such that for every $\omega$, $\tilde\sigma^{\tau,\omega}=\sigma$. Then, we have for $\mathbb P$-a.e. $\omega$
\begin{align*}
\E^{\P^{\tau,\omega}}[\abs{B_{\sigma}^{\tau}}] &\leq\E^{\P^{\tau,\omega}}[\abs{B^{\tau,\omega}_{\tilde{\sigma}^{\tau,\omega}}}]+\abs{B_{\tau}(\omega)}= \E_{\tau}^{\mathbb{P}}[\abs{B_{\tilde{\sigma}}}](\omega) + \abs{B_{\tau}}(\omega)<+\infty,
\end{align*}
where we used the fact that $\mathbb P\in{\mathcal P}_m$.  Similarly, we have for $\mathbb P$-a.e. $\omega$:
\begin{align*}
\E^{\P^{\tau,\omega}}[B_{\sigma}^{\tau}] &=\E^{\P^{\tau,\omega}}[B^{\tau,\omega}_{\tilde{\sigma}^{\tau,\omega}}-B_{\tau}(\omega)]= \E_{\tau}^{\mathbb{P}}[B_{\tilde{\sigma}}](\omega) - B_{\tau}(\omega)=0.
\end{align*}
By the arbitrariness of $\tau$ and $\sigma$, this completes the verification of Condition \ref{cond.nutz} (ii).

\vspace{0.3em}
To verify Condition \ref{cond.nutz} (ii), we adapt an argument from \cite{nn}. We define the following map
$$ \psi : \mathbb{H}^1_{\text{loc}}(\P_0,\mathbb{S}_d^{>0},\F) \rightarrow \mathbf{M}(\Omega), \ \ \ \alpha \mapsto \P^{\alpha}=\P_0\circ \left( \int_0^{\cdot}\alpha_s^{1/2}dB_s \right)^{-1}. $$

From \cite{nn} (see Lemmas 3.1 and 3.2), we know that it is sufficient to show that $ \mathcal{P}_m \subset \mathbf{M}(\Omega) $ is the image of a Borel space (i.e. a Borel subset of a Polish space) under a Borel map. For that we show that $ \mathbb{H}^0(\mathbb{S}^{>0}_d,\F) $ is Polish and $ \mathbb{H}^1_{\text{loc}}(\P_0,\mathbb{S}_d^{>0},\F)\subset \mathbb{H}^0(\mathbb{S}^{>0}_d,\F) $ is Borel. The first part is already given in Lemma 3.1 of \cite{nn}. We then need to show that the map $ \psi : \mathbb{H}^1_{\text{loc}}(\P_0,\mathbb{S}_d^{>0},\F) \rightarrow \mathbf{M}(\Omega) $ is Borel, which is a direct consequence of Lemma 3.2 in \cite{nn}.

\vspace{0.3em}
It then only remains to prove that $ \mathbb{H}^1_{\text{m}}(\P_0,\mathbb{S}_d^{>0},\F) \subset \mathbb{H}^0(\mathbb{S}_d^{>0},\F) \ \  \text{is Borel}, $where
$$ \mathbb{H}^1_{\text{m}}(\P_0,\mathbb{S}_d^{>0},\F):=\left\{ \alpha \in  \mathbb{H}^0(\mathbb{S}^{>0}_d,\F) \ : \ \sup_{\tau} \E^{\mathbb P_0}\left[|X^{\alpha}_{\tau}|{\bf{1}}_{|X^{\alpha}_{\tau}|\geq n}\right]  \underset{n \rightarrow + \infty}{\longrightarrow}  0 \right\}. $$

It is clear that
$$ \mathbb{H}^1_{\text{m}}(\P_0,\mathbb{S}_d^{>0},\F)= \underset{p\in \mathbb{N}^*}{\bigcap} \underset{N \in \mathbb{N}}{\bigcup} \ \underset{n \geq N}{\bigcap} \left\{ \alpha \in \mathbb{H}^0(\mathbb{S}_d^{>0},\F) \ : \ \sup_{\tau} \E^{\P_0}[|X^{\alpha}_{\tau}|{\bf{1}}_{|X^{\alpha}_{\tau}|> n}] \le \frac{1}{p} \right\}.  $$
and
$$ \left\{ \alpha \in \mathbb{H}^0(\mathbb{S}^{>0}_d,\F): \ \sup_{\tau} \E^{\P_0}[|X^{\alpha}_{\tau}|{\bf{1}}_{|X^{\alpha}_{\tau}|> n}] \le \frac{1}{p} \right\}=\psi^{-1} \left\{ \P \in \mathcal{P}_S : {\scriptstyle \ \sup_{\tau} \E^{\P_0}[|B_{\tau}|{\bf{1}}_{|B_{\tau}| >n}] \le \frac{1}{p} }\right\}. $$
It then suffices to show that for any $ n \in  \mathbb{N} $, the following function $f_n$ is Borel measurable:
$$ f_n \ : \ \P \mapsto \sup_{\tau} \E^{\P}[|B_{\tau}|{\bf{1}}_{|B_{\tau}| >n}]. $$
We actually show that this function is lower semi-continuous. For $K,l>0$, define:
 \b*
 \phi(x)=|x|{\bf{1}}_{|x| >n}
 &\mbox{and}&
 \phi_{K,l}(x)=|x| \wedge K \frac{(|x|-n)^+-(|x|-n-l)^+}{l},
 ~~x\in\R^d.
 \e*
We emphasize that $ \phi_{K,l} $ is uniformly continuous and bounded. Let then $\mathbb P\in\mathcal P_{S}$ and consider some sequence $ (\P^i)_{i\geq 0}$ which converges weakly to $\mathbb P$. We represent $ \P^i $ and $ \P $ by $ \alpha_i $ and $ \alpha$. Remember that for any $\widetilde \P\in\mathcal P_S$, associated to some $\widetilde\alpha$, we have
$$ f_n(\widetilde{\P})= \sup_{\tau} \E^{\mathbb P_0}\left[|X^{\widetilde\alpha}_{\tau}|{\bf{1}}_{|X^{\widetilde\alpha}_{\tau}|> n}\right].$$
The weak convergence of $\mathbb P^i$ to $\mathbb P$ is equivalent to the convergence in law of $X^{\alpha_i}$ to $X^\alpha$. Hence, for all $ \tau $, we have
$$ \underset{ i \rightarrow +\infty}{\lim} \E^{\P_0}  [ \phi_{K,l}(X_{\tau}^{\alpha_i})]= \E^{\P_0}  [ \phi_{K,l}(X_{\tau}^{\alpha})], $$
from which we deduce easily $ \underset{ i \rightarrow +\infty}{\liminf} \sup_{\tau} \E^{\P_0} [ \phi_{K,l}(X_{\tau}^{\alpha_i})] \geq \E^{\P_0} [ \phi_{K,l}(X_{\tau}^{\alpha})].$ As this is true for all $ K>0$ and $ l>0$, and since the function $\phi_{K,l}$ is non-decreasing in $K$ and non-increasing in $l$, we also have 
$$ \underset{ i \rightarrow +\infty}{\liminf} \sup_{\tau} \E^{\P_0}  [ \phi(X_{\tau}^{\alpha_i})] \geq \E^{\P_0}  [ \phi_{K,l}(X_{\tau}^{\alpha})].  $$
Letting $K$ go to $+\infty$ and $l$ to $0$ on the right-hand side above, we deduce using monotone convergence and taking supremum in $\tau$
$$ \underset{ i \rightarrow +\infty}{\liminf} \sup_{\tau} \E^{\P_0}  [ \phi(X_{\tau}^{\alpha_i})] \geq  \sup_{\tau} \E^{\P_0}  [ \phi(X_{\tau}^{\alpha})] $$
Then $ f_n $ is lower semicontinuous and thus measurable.
\begin{flushright}$\Box$\end{flushright}

\section{The duality result}
\label{sect:duality}
In this section we show our main result Theorem \ref{th.main}. We will assume throughout that $\xi$ is upper semi-analytic. For that purpose, we introduce the dynamic version of the dual problem:
$$ 
Y_t(\xi)(\omega) \mathrel{\mathop:}= \sup_{\mathbb{P} \in \mathcal P_m(t,\omega)} \E^{\mathbb{P}} [ \xi^{t,\omega}],
~~t\in[0,T],~\omega\in\Omega.
$$
We first observe that $Y_t$ is measurable with respect to the universal filtration $\mathcal{F}^*_t $, as a consequence of Step $1$ in the proof of Theorem 2.3 in \cite{nvh}, since Condition \ref{cond.nutz}(i) holds true for $\mathcal P_m$.

\begin{Lemma} \label{essup}
Let $\tau$ be an $\mathbb F$-stopping time. Then, for all $\P \in \mathcal{P}_m$:
 \b*
 Y_{\tau}(\xi)
 &=&
 \underset{\P' \in \mathcal{P}_m(\tau,\P)}
 {\esup^{\P}} \E^{\P'}[\xi | \mathcal{F}_{\tau}], 
 ~~\P-\mbox{a.s.}
 \e*
where $ \mathcal{P}_m(\tau,\P)=\lbrace \P' \in \mathcal{P}_m \ : \ \P'=\P \ \text{on} \ \mathcal{F}_{\tau} \rbrace$.
\end{Lemma}

\vspace{0.3em}
\textbf{Proof}
The inequality $ \geq $ is trivial as $ Y_\tau $ is $\mathcal{F}^*_{\tau}$-measurable and measures extend uniquely to universal completions, which means that if $\P$ and $\P'$ coincide on $\mathcal F_\tau$, they also coincide on $\mathcal F^*_\tau$ (see Step 3 of the proof of Theorem $2.3$ in \cite{nvh} for similar arguments). We then focus on $ \le $. Fix some $ \mathbb{P} \in \mathcal{P}_m $. We recall that following the same construction as in Step 2 of the proof of Theorem 2.3 in \cite{nvh}, for any $ \epsilon>0$, we can construct a $ \mathcal{F}_{\tau}$-measurable kernel $\nu:\Omega \rightarrow \mathbf{M}(\Omega)$ such that:
\begin{equation}\label{eps} \E^{\nu(\omega)}[\xi^{\tau,\omega}] 
\geq 
\big(Y_{\tau}(\omega)-\epsilon\big)
\1_{\{-\infty<Y_{\tau}(\omega)<\infty\}}
+\epsilon^{-1}\1_{\{Y_{\tau}(\omega)=\infty\}}
-\infty\1_{\{Y_\tau(\omega)=-\infty\}},
\end{equation}
and such that $\nu(\omega) \in \mathcal{P}_m(\tau,\omega) $, $ \P$-a.s.

\vspace{0.3em}
We then consider the probability $ \widetilde{\P} \in \mathcal{P}_S(\tau,\P) $ associated to $\nu$ through Condition \ref{cond.nutz}(iii), where this last assertion uses that Condition \ref{cond.nutz}(iii) is verified for $ \mathcal{P}_S$ thanks to Theorem 2.4 in \cite{nn}. However there is no guarantee that  $\widetilde{\P}$ belongs to $\mathcal P_m$, and the rest of this proof overcomes this difficulty by using a suitable approximation.

\vspace{0.3em}
{\bf Step 1:} Construction of an approximation $ \nu^n$. Define, for all $n\geq 1$,
 \b*
 \nu^n(\omega) 
 &:=& 
 \nu(\omega)
 \1_{\{\E^{\nu(\omega)}[|B_T^{\tau}|] \le n\}}
 +\P^{\tau,\omega}
 \1_{\{\E^{\nu(\omega)}[|B_T^{\tau}|] > n\}}.
 \e*
Clearly, $\nu^n$ is a measurable kernel, and since $\mathcal P_m$ is stable by bifurcation, we also have $\nu^n(\omega)\in\mathcal P_m(\tau,\omega)$. Observe that $ E_n \mathrel{\mathop:}= \lbrace \omega \in \Omega \ : \ \E^{\nu(\omega)}[|B_T^{\tau}|] \le n \rbrace,
~n\ge 1, $
is an increasing sequence in $\mathcal F_\tau$ with $\P(E_n)\underset{n \rightarrow + \infty}{ \longrightarrow}1 $, as a consequence of the fact that $ \E^{\nu(\omega)}[|B_T^{\tau}|] <+\infty,$ $\P-$a.s.
We now define the measure $\overline{\P}^n$ by:
$$\overline{\P}^n(A) \mathrel{\mathop:}= \iint ({\bf{1}}_A)^{\tau,\omega} ( \omega') \nu^n(d \omega' ; \omega ) \mathbb{P}(dw), \ \ A \in \mathcal{F}.$$

We first show that $ \overline{\P}^n \in \mathcal{P}_m(\tau,\P)$. The fact that $\overline\P^n$ coincides with $\mathbb P$ on $\mathcal F_\tau$ is clear by construction. Next, we compute that:
\begin{align*}
\E^{\overline{\P}^n}[|B_T|] &= \E^{\overline{\P}^n}\left[|B_T|{\bf{1}}_{E_n} +|B_T|{\bf{1}}_{E_n^c} \right ] \\
& = \E^{\overline{\P}^n}\left[\mathbb E^{\nu^n(\omega)}\left[|B_T^{\tau,\omega}|\right]{\bf{1}}_{E_n}\right] + \E^{\overline\P^n}\left[\mathbb E^{\nu^n(\omega)}\left[|B_T^{\tau,\omega}|\right]{\bf{1}}_{E_n^c}\right]   \\
& = \E^{\overline{\P}^n}\left[\mathbb E^{\nu(\omega)}\left[|B_T^{\tau,\omega}|\right]{\bf{1}}_{E_n}\right] + \E^{\overline\P^n}\left[\mathbb E^{\mathbb P^{\tau,\omega}}\left[|B_T^{\tau,\omega}|\right]{\bf{1}}_{E_n^c}\right]   \\
& \leq \E^{\P}\left[\left(\mathbb E^{\nu(\omega)}\left[|B_T^{\tau}|\right]+\abs{B_\tau}\right){\bf{1}}_{E_n}\right] + \E^{\P}\left[\mathbb E^{\mathbb P^{\tau,\omega}}\left[|B_T^{\tau,\omega}|\right]{\bf{1}}_{E_n^c}\right]\\
 & \le \mathbb E^{\P}\left[\abs{B_\tau}\right]+n+\mathbb E^\mathbb P\left[\abs{B_T}\right]<+\infty. 
\end{align*}
To prove the martingale property of $B$ under $\overline{\P}^n$, we consider an arbitrary $\mathbb F$-stopping time $\sigma$, and we compute that:
 $$ 
 \E^{\overline{\P}^n}[B_{\sigma}] 
 = 
 \E^{\overline{\P}^n}[B_{\sigma}  {\bf{1}}_{\sigma \le \tau} 
                 + B_{\sigma} {\bf{1}}_{\sigma > \tau} ] 
 = 
 \E^{\P}[B_{\sigma}  {\bf{1}}_{\sigma \le \tau}]
 +\E^{\overline{\P}^n}[B_{\sigma} {\bf{1}}_{\sigma > \tau} ], 
 $$
by the fact that $\overline\P^n=\P$ on $\mathcal F_\tau$. We continue computing
\begin{align*}
\E^{\overline{\P}^n}[ B_{\sigma} {\bf{1}}_{\sigma > \tau} ] = \E^{\overline{\P}^n} \left[ \E^{(\overline{\P}^{n})^{\tau,\omega}}\left[B_{\sigma^{\tau,\omega}}^{\tau,\omega}\right] {\bf{1}}_{\sigma > \tau}\right] &= \E^{\overline{\P}^n} \left[ \E^{\nu^n(\omega)} \left[B^{\tau}_{\sigma^{\tau,\omega}}+B_{\tau}(\omega)\right]  {\bf{1}}_{\sigma > \tau}\right] \\
&=\E^{\overline{\P}^n} \left[B_{\tau}{\bf{1}}_{\sigma > \tau}\right]
=\E^{\P} \left[B_{\tau}{\bf{1}}_{\sigma > \tau}\right],
\end{align*}
where the last equality uses the definition of $ \nu^n $ which ensures that $ \nu^n(\omega) \in \mathcal{P}_m(\tau,\omega)$, $ \P$-a.s. Therefore, we have $\E^{\overline{\P}^n}[B_{\sigma}]=\mathbb E^\mathbb P[B_{\sigma\wedge\tau}]=0,$
since $B$ is a martingale under $\mathbb P$.

\vspace{0.3em}
{\bf Step 2:} By \reff{eps}, we have for every $\omega$
$$ \E^{\nu^n(\omega)}[\xi^{\tau,\omega}] \geq (Y_{\tau}(\omega)-\epsilon)\wedge \eps^{-1} {\bf{1}}_{E_n} + \E^{\P^{\tau,\omega}}[\xi^{\tau,\omega}] {\bf{1}}_{E_n^c}. $$
Then for any $\omega\in\Omega\backslash\mathcal N^\mathbb P$, for some $\mathbb P$-null set $\mathcal N^\mathbb P$
$$ \E^{\overline{\P}^n}[\xi | \mathcal{F}_{\tau}](\omega) \geq (Y_{\tau}(\omega)-\epsilon)\wedge\eps^{-1}{\bf{1}}_{E_n}(\omega)+\E^{\P}[\xi| \mathcal{F}_{\tau} ](\omega){\bf{1}}_{E_n^c}(\omega).$$

\vspace{0.3em}
Hence, for any $\omega\in\Omega\backslash\mathcal N^\mathbb P$, for all $ n \geq 0$
\begin{align*}
\underset{\P' \in \mathcal{P}_m(\tau,\P)}{\esup^{\P}} \E^{\P'}[\xi | \mathcal{F}_{\tau}](\omega) & \geq  (Y_{\tau}(\omega)-\epsilon)\wedge\eps^{-1}{\bf{1}}_{E_n}(\omega)+\E^{\P}[\xi| \mathcal{F}_{\tau} ](\omega){\bf{1}}_{E_n^c}(\omega).
\end{align*}
We emphasize that \textit{a priori}, the right-hand side above is only $\mathcal F_\tau^*$-measurable. However, if $\mathbb P$ and $\mathbb P'$ coincide on $\mathcal F_\tau$, they also coincide on $\mathcal F_\tau^*$, since measures extend uniquely on universal completions. Therefore the above inequality does indeed hold $\mathbb P-a.s.$

\vspace{0.3em}
Since the sequence $E_n$ increases to $\Omega$ (up to some $\mathbb P$-null set which we implicitly add to $\mathcal N^\P$), for any $\omega\in\Omega\backslash\mathcal N^\mathbb P$, there exists $N(\omega)\in\mathbb N$ such that if $n\geq N(\omega)$, then $\omega\in E_n$. Therefore, taking $n$ large enough, we have
\begin{align}\label{eq1}
\underset{\P' \in \mathcal{P}_m(\tau,\P)}{\esup^{\P}} \E^{\P'}[\xi | \mathcal{F}_{\tau}](\omega) & \geq  (Y_{\tau}(\omega)-\epsilon)\wedge\eps^{-1}.
\end{align}

If $Y_\tau(\omega)=-\infty$, then by the inequality proved at the beginning, the left-hand side above is also equal to $-\infty$. Hence the result in this case. If $Y_\tau(\omega)=+\infty$, then \reff{eq1} implies directly that the left-hand side is also $+\infty$ by arbitrariness of $\eps>0$. Finally, if $Y_\tau(\omega)$ is finite, the desired result follows from \reff{eq1} by arbitrariness of $\eps$.
\begin{flushright}$\Box$\end{flushright}

We then continue with a version of the tower property in our context.

\begin{Proposition} \label{tp}
Let $\P \in \mathcal{P}_m^\xi$, and $\sigma,\tau$ two $\mathbb{F}$-stopping times with $\sigma\le\tau$. Then, $\P-$a.s.
 $$
 Y_{\sigma}(\xi) 
 =\!\!\!\!
 \underset{\P'\in \mathcal{P}^\xi_m(\sigma,\P)}{\esup}\E^{\P'}\!\!
 \left[ \underset{\P''\in \mathcal{P}_m^\xi(\tau,\P')}{\esup}\E^{\P''}[ \xi | \mathcal{F}_{\tau}]   |\mathcal{F}_{\sigma}
 \right]  
 = \!\!\!\underset{\P'\in \mathcal{P}_m(\sigma,\P)}{\esup}\E^{\P'}\!\!
 \left[ \underset{\P''\in \mathcal{P}_m(\tau,\P')}{\esup}\E^{\P''}[ \xi | \mathcal{F}_{\tau}]   |\mathcal{F}_{\sigma}
 \right],
 $$
where for any $\mathbb F$-stopping time $\iota$ and any $\mathbb P\in\mathcal P^\xi_m$
$$\mathcal{P}_m^\xi(\iota,\P):=\left\{\mathbb P'\in\mathcal P_m^\xi\text{ s.t. }\mathbb P'=\mathbb P \text{ on }\mathcal F_\iota\right\}.$$
\end{Proposition}

\textbf{Proof}
We consider $ \P \in \mathcal{P}_m^\xi $. Exactly as in the proof of Lemma \ref{essup}, we can construct a measurable kernel $ \nu^n $ from a kernel $ \nu$ such that:
\\
$\bullet$ $ \nu^n $ is $ \mathcal{F}_{\tau}$-measurable.
\\
$\bullet$ $ \P^n \in \mathcal{P}_m(\tau,\P)$ where $ \P^n(A)= \iint ({\bf{1}}_A)^{\tau,\omega} ( \omega')\nu^n(d \omega' ; \omega ) \mathbb{P}(dw), \ \ A \in \mathcal{F}$.
\\
$\bullet$ $ \nu $ is a $ \mathcal{F}_{\tau}$-measurable kernel such that \reff{eps} holds.
\\
$\bullet$ $ E_n= \lbrace{ \nu^n=\nu}\rbrace$ is an increasing sequence such that $ \P(E_n) \underset{n \rightarrow + \infty}{\longrightarrow}1$.

We then compute for any $\eps>0$
 \b*
 \underset{\P' \in \mathcal{P}_m(\sigma,\P)}{\esup}
 \E^{\P'}[\xi | \mathcal{F}_{\sigma}]
 \;\ge\;
 \E^{\P^n}[\xi | \mathcal{F}_{\tau}]
 &\ge&
 \E^{\P}[(Y_{\tau}-\epsilon)\wedge\epsilon^{-1}{\bf{1}}_{E_n} - \E^\P[\xi^-| \mathcal{F}_{\tau}] {\bf{1}}_{E_n^c} | \mathcal{F}_{\sigma}],\ \mathbb P-a.s.
 \e* 
Recall that $\E^{\P}[\xi^-]<\infty$. Then, it follows from the dominated convergence theorem that $\E^{\P}[\xi^- {\bf{1}}_{E_n^c} | \mathcal{F}_{\sigma}]\longrightarrow 0$, as $n \to\infty$, $\P-$a.s. Also, since $Y_{\tau}\ge-\E^\P[\xi^-|{\cal F}_\tau]\in\L^1(\P)$, it follows from Fatou's lemma that:
$$ \E^{\P}[(Y_{\tau}-\epsilon)\wedge\epsilon^{-1} | \mathcal{F}_{\sigma}] \le \underset{\P' \in \mathcal{P}_m(\sigma,\P)}{\esup}\E^{\P'}[\xi | \mathcal{F}_{\sigma}]  \ \P-\text{a.s.}$$
Finally, when $ \epsilon \rightarrow 0$, we have $ \P$-a.s., with the last equality being obvious
$$
\E^{\P}[Y_{\tau}| \mathcal{F}_{\sigma}] 
\le 
\underset{\P' \in \mathcal{P}_m(\sigma,\P)}{\esup}\E^{\P'}[\xi | \mathcal{F}_{\sigma}]
=\underset{\P' \in \mathcal{P}_m^\xi(\sigma,\P)}{\esup}\E^{\P'}[\xi | \mathcal{F}_{\sigma}].
$$ \begin{flushright}$\Box$\end{flushright}

\begin{Proposition} \label{supermart}
Assume that $\sup_{\P\in\mathcal{P}_m}\E^{\P}[\xi^+]<\infty $. Then, for any $ \P \in \mathcal{P}^\xi_m$, the process $\{Y_t(\xi),t\le T\}$ is a $\P$-supermartingale.
\end{Proposition}

\textbf{Proof}
In view of Proposition \ref{tp}, we already have the tower property. It only remains to show the integrability of $ Y_t(\xi)$ for all $ t\in [0,T]$. For that we only need to show that $ Y_t(\xi^+) $ is integrable. We fix $ 0 \le t  \le T$ and $ \mathbb{P} \in \mathcal{P}^\xi_m$. Then $ Y_t(\xi^+)(\omega) \in \mathbb{R}_+ \times \lbrace + \infty \rbrace $. Let us then show that the family $ \lbrace \E^{\mathbb{P}'}[ \xi^+ |\mathcal{F}_t ], \ \mathbb{P}'\in \mathcal{P}_m(t,\mathbb{P})  \rbrace $ is upward directed.

\vspace{0.3em}
We consider $ \mathbb{P}_1 $ and $ \mathbb{P}_2 $ in $ \mathcal{P}_m(s,\mathbb{P})  $. The set $ A \mathrel{\mathop:}= \lbrace \mathbb{E}^{\mathbb{P}_2}[\xi^+ | \mathcal{F}_t] \le  \mathbb{E}^{\mathbb{P}_1}[\xi^+ | \mathcal{F}_t] \rbrace, $
is $ \mathcal{F}_t $-measurable. Then $ \overline{\mathbb{P}} \mathrel{\mathop:}= \mathbb{P}_1 {\bf{1}}_{A} + \mathbb{P}_2 {\bf{1}}_{A^c} $ is an element of $  \mathcal{P}_m(t,\mathbb{P}) $ such that:
$$ \E^{\overline{\mathbb{P}}} [\xi^+ | \mathcal{F}_t ] = \mathbb{E}^{\mathbb{P}_1}[\xi^+ | \mathcal{F}_t] \vee \mathbb{E}^{\mathbb{P}_2}[\xi^+ | \mathcal{F}_t]  $$
We then have an increasing sequence $ \mathbb{P}_n $ of $ \mathcal{P}_m(t,\mathbb{P}) $ such that:

$$ \mathbb{E}^{\mathbb{P}_n}[\xi^+ | \mathcal{F}_t] \nearrow \underset{\mathbb{P}' \in \mathcal{P}_m(s,\mathbb{P})}{\esup} \E^{\mathbb{P}'} [\xi^+ | \mathcal{F}_t] , \  \mathbb{P}-\text{a.s.},  $$
and by the monotone convergence theorem $\lim_{n \rightarrow + \infty} \E^{\mathbb{P}_n} [\xi^+] = \E^{\mathbb{P}} [Y_t (\xi^+)].$ Hence
$$  \E^{\mathbb{P}} [Y_t (\xi^+)] \le \sup_{\mathbb{P} \in \mathcal{P}_m} \E^{\mathbb{P}} [\xi^+ ] <+\infty,\ \text{ for all } \mathbb{P} \in \mathcal{P}_m. $$
\begin{flushright}$\Box$\end{flushright}

We now have all ingredients to follow the classical line of argument for the

\vspace{0.3em}
\no {\bf Proof of theorem \ref{th.main}}
For the sake of simplicity, the dependence of $Y$ in $\xi$ will be omitted.

\vspace{0.3em}
(i) We first show that right-limiting process $ \overline{Y}_t \mathrel{\mathop:}= Y_{t+}$, $t\le T,$ is a $ (\mathbb{G}^{\mathcal P_m}, \mathbb{P})$ supermartingale for all $\P\in{\cal P}_m^\xi$.
By Proposition \ref{supermart} and the fact that for any $ \mathbb{P} \in \mathcal{P}_{S}, $ $ \overline{\mathbb{F}}^{\mathbb{P}} $ is right-continuous, contains $ \mathbb{G} $ and $ \mathbb{P}$ has the predictable representation property (see \cite{stz-quasi}), $ Y $ is a $ (\mathbb{F}^*,\mathbb{P})$ supermartingale for every $ \mathbb{P} \in \mathcal{P}_m^\xi $. Then applying \cite{dm} (see Theorem YI.2 page 67), we have that $ \overline{Y} $ is well defined $ \mathbb{P}$-a.s. and $ \overline{Y}$ is a right continuous $ (\mathbb{G}^{\mathcal P_m},\mathbb{P})$ supermartingale for all $ \mathbb{P} \in \mathcal{P}_m^\xi .$ We also notice the important fact that for all $ \mathbb{P} \in \mathcal{P}_m^\xi $ we have $ \overline{Y}_t \le Y_t $ $ \mathbb{P}$-a.s. In particular, $ \overline{Y}_0 \le Y_0, $ and $ \overline{Y}_0 $ is constant because $ \mathcal{G}_0^{\mathcal P_m} $ is trivial.

\vspace{0.3em}
(ii) We next construct the optimal trading strategy.
By the Doob-Meyer decomposition (see Theorem 13 page 115 in \cite{prot}), there exists a pair of processes $ (H^{\mathbb{P}},K^{\mathbb{P}} )$ where $ H^{\mathbb{P}}$ belongs to $\mathbb{H}^2_{\text{loc}}(\mathbb{P},\R^d,\G^{\mathcal P_m})$ and $ K^{\mathbb{P}} $ is $ \mathbb{P}-$integrable and non-decreasing, such that:
$$ \overline{Y}_t=Y_0+\int_0^t H_s^{\mathbb{P}} dB_s - K_t^{\mathbb{P}}, \ t\in [0,T], \ \mathbb{P}- \text{a.s.} $$ 
Since $ \overline{Y} $ is right continuous, it follows from Karandikar \cite{k} that the family $ H^{\mathbb{P}} $ can be aggregated by some process $\widehat H$ in the sense that $ \widehat{H}=H^{\mathbb{P}} $, $ dt \times d\mathbb{P}$-a.s. for all $ \mathbb{P}\in \mathcal{P}_m^\xi$. Thus, for every $ \mathbb{P} \in \mathcal{P}_m^\xi $, the local martingale $ \int \widehat{H} dB $ is bounded from below by the martingale $ \E^{\mathbb{P}}[\xi | \mathcal{G}_t^{\mathcal P_m}] $. Hence this is a supermartingale which ensures that $ \widehat{H} $ is in $ \mathcal{H}^\xi$ and superreplicates the claim $\xi$ $\mathcal P^\xi_m$-quasi-surely.
\begin{flushright}$\Box$\end{flushright}

\section{Extensions}
\label{sect:extensions}
\subsection{The case of $ \mathcal{P}_b$}

In this section, we show that Theorem \ref{th.main} together with the previous arguments in its proof, hold for the following example, which is important in the context of second-order BSDEs as introduced in \cite{stz}. We recall that $<B>$ is well defined pathwise and that its density is denoted by $\widehat a$.
\begin{Definition}
$ \mathcal{P}_b $ is the sub-class of $ \mathcal{P}_S $ consisting of all $ \mathbb{P} \in \mathcal{P}_S $ such that:
$$ \underline{a}_{\P} \le \widehat{a} \le \overline{a}_{\P}, \ \ dt \times d\P- \text{a.s. for some} \ \underline{a}_{\P}, \overline{a}_\P \in \mathbb{S}^{>0}_d. $$
\end{Definition}

We emphasize that our Example \ref{example} also shows directly that $\mathcal P_b$ does not satisfy Condition \ref{cond.nutz}(iii). We now prove the three main technical results of this paper in this case.

\begin{Proposition}
$ \mathcal{P}_b $ verifies Condition \ref{cond.nutz} (i) and (ii).
\end{Proposition}

\textbf{Proof}
(ii) has already been proved in \cite{stz-dual}. Let us now prove (i). We observe that:
$$ \mathcal{P}_b=\underset{\underline{a},\overline{a}\in \mathbb{S}_d^{>0}(\mathbb{Q})}{\bigcup} \lbrace \P \in \mathcal{P}_S \ : \ \underline{a} \le \hat{a} \le \bar{a} \ dt \times d\P- \text{a.s.} \rbrace. $$
By Proposition 3.1 in \cite{nvh}, we know that all these sets satisfy Condition \ref{cond.nutz}(i). Since a countable union of analytic set is analytic, we obtain the result. \begin{flushright}$\Box$\end{flushright}

It remains to introduce a suitable sequence of approximations of measurable kernels as in the proof of Lemma \ref{essup} and Proposition \ref{tp}.

\begin{Proposition}\label{pb}
The results of Lemma \ref{essup}, Proposition \ref{tp} and \ref{supermart} and Theorem \ref{th.main} are valid if we replace $ \mathcal{P}_m$ by $ \mathcal{P}_b$.
\end{Proposition}

\textbf{Proof} We only redefine an approximated kernel family adapted to $ \mathcal{P}_b$, which allows, by the same arguments as in Lemma \ref{essup} and Proposition \ref{supermart}, to obtain the duality result of Theorem \ref{th.main}. Let $ \tau $ be a $ \mathbb{F}$-stopping time and $\nu$ the $ \mathcal{F}_{\tau}$-measurable kernel obtained by the same construction as in Lemma \ref{essup} and Proposition \ref{tp}. For $ \P\in \mathcal{P}_b$ we are interested in the measure $ \overline{\P}$ defined by:
\begin{equation*}
\overline{\mathbb P}(A)= \iint \left({\bf{1}}_A\right)^{\tau,\omega} \left( \omega'\right) \nu\left(d \omega' ; \omega \right) \mathbb{P}\left(dw\right).
\end{equation*}
 
Then $ \overline{\P}$ is in $ \cal P_S$ and there is some $\alpha$ s.t. $ \overline{\P}= \mathbb{P}_0 \circ (X^{\alpha}_.)^{-1}$. Define $ \overline{\P}^n$ by:
$$  \overline{\P}^n:= \P_0 \circ \left( \int_0^{\cdot}(\alpha_s^{1/2}{\bf{1}}_{s \le \tau} + \pi_n(\alpha_s^{1/2}){\bf{1}}_{s> \tau} )dB_s) \right)^{-1},\ n\geq 1, $$
where $ \pi_n $ is the projection on $\overline{\mathcal B}_{\mathbb S^{>0}_d}(0,n)\backslash \overline{\mathcal B}_{\mathbb S^{>0}_d}(0,1/n)$, where $\overline{\mathcal B}_{\mathbb S^{>0}_d}(x,r)$ denotes the closed ball of $\mathbb S^{>0}_d$ centered at $x$ with radius $r$. Then $\overline{\P}^n$ belongs to $\mathcal P_b$. Observe also that the sets $$ E_n:= \left\{\omega\in\Omega,\  \left(\overline{\P}^{n}\right)^{\tau,\omega}=\nu(\omega)\right\}, $$ are in $\mathcal F_\tau$ and define an increasing covering of $\Omega$. We then build the "right" approximation $ \widetilde{\P}^n$, ensuring all the convergences in the proofs, by $ \widetilde{\P}^n= \overline{\P}^n {\bf{1}}_{E_n}+\P {\bf{1}}_{E_n^c} $.
$ \widetilde{\P}^n $ is in $ \mathcal{P}_b$, and associated to the $ \mathcal{F}_{\tau}$-measurable kernel $ \tilde{\nu}^n(\omega):=\nu(\omega){\bf{1}}_{\omega \in E_n}+\P^{\tau,\omega}{\bf{1}}_{\omega \in E_n^c} $.
 Then we can reproduce exactly the same proofs as in the case of $\mathcal P_m$.
\begin{flushright}$\Box$\end{flushright}

\subsection{A general framework}
As the reader may have realized, our proofs in the case of $\mathcal P_m$ and $\mathcal P_b$ are very similar, and essentially rely on the construction of a suitable approximated kernel. In this subsection, we consider a generic subset $\mathcal P$ of $\mathcal P_S$ (and the corresponding shifted families $\mathcal P(s,\omega)$), and we give a general condition, (weaker than Condition \ref{cond.nutz}) under which our results still hold true. We recall that such a family is said to be stable by bifurcation if for any $\F$-stopping times $ 0 \le \sigma \le \tau$, $ \omega\in\Omega$, $ A$ $ \mathcal{F}_{\tau}$-measurable, $ \P_1$ and $ \P_2$ in $ \mathcal{P}(\sigma,\omega)$, we have $$ \P=\P_1 {\bf{1}}_A + \P_2 {\bf{1}}_{A^c} \in \mathcal{P}(\sigma,\omega).$$

\begin{Condition} \label{cond.nutz'}
Let $s\in \mathbb{R}_+$, $ \tau\geq s$ a stopping time, $\overline{\omega}\in \Omega$, $ \mathbb{P}\in \mathcal{P}(s,\overline{\omega})$ and $ \theta \mathrel{\mathop:}=\tau^{s,\overline{\omega}}-s $.
\\
{\rm{(i)}} The graph $ \lbrace (\mathbb{P}',\omega): \omega \in \Omega, \ \mathbb{P}' \in \mathcal{P}(t,\omega) \rbrace \subseteq \mathbf{M}(\Omega) \times \Omega $ is analytic.
\\
{\rm{(ii)}}  We have $ \mathbb{P}^{\theta,\omega} \in \mathcal{P}(\tau,\overline{\omega} \otimes_s \omega ) $ for $ \mathbb{P}$-a.e. $ \omega \in \Omega $.
\\
{\rm{(iii)}}  $ \mathcal{P}$ is stable by bifurcation.
\\
{\rm{(iv)}} If $ \nu : \Omega \rightarrow \mathbf{M}(\Omega)$ is an $ \mathcal{F}_{\theta}$-measurable kernel and $ \nu(\omega) \in \mathcal{P}(\tau,\overline{\omega}\otimes_s \omega)$ for $ \mathbb{P}$-a.e. $ \omega \in \Omega $, then there exists $ \nu^n : \Omega \rightarrow \mathbf{M}(\Omega)$, which is a $ \mathcal{F}_{\theta}$-measurable kernel such that $ \P(\nu^n=\nu)\underset{n\rightarrow \infty}{\longrightarrow} 1$ and the following measure $\overline{\mathbb P}^n\in \mathcal{P}(s,\overline{\omega})$:
$$\overline{\mathbb P}^n(A)= \iint ({\bf{1}}_A)^{\theta,\omega} ( \omega') \nu^n(d \omega' ; \omega ) \mathbb{P}(dw), \ \ A \in \mathcal{F}.$$
\end{Condition}

\begin{Remark}
Notice that Condition \ref{cond.nutz'} is weaker than Condition \ref{cond.nutz}. Indeed, Condition \ref{cond.nutz}(iii) implies directly that the set $\mathcal P$ is stable by bifurcation. Moreover, considering the constant kernels $\nu^n:=\nu$, it also implies Condition \ref{cond.nutz'}(iv). Furthermore, as shown in our previous proofs, the sets $\mathcal P_m$ and $\mathcal P_b$ satisfy Condition \ref{cond.nutz'} but not Condition \ref{cond.nutz}.
\end{Remark}

Similarly to our previous notations, we introduce the sets $ \mathcal{H}^\xi(\mathcal{P})$ and $ \mathcal{P}^\xi$. In this context, we obtain a new version of Theorem \ref{th.main}:

\begin{Theorem}\label{th.main.ext}
Let $\mathcal P(s,\omega)$ be a family of probability measures satisfying Condition \ref{cond.nutz'}. Let $\xi$ be an upper semi-analytic r.v. with $\sup_{\P\in\mathcal{P}}\E^{\P}[\xi^+]<+\infty $. Then
 \b*
 V(\xi) 
 &\mathrel{\mathop:}=& 
 \inf \left\{X_0:~X^H_T \ge \xi, \ \ \mathcal P^\xi-
                 \mbox{q.s. for some}~H\in{\cal H}^\xi(\mathcal P) 
      \right\} 
 \;=\; 
 \sup_{\P \in \mathcal{P}} \E^{\P}[\xi].
 \e*
Moreover, existence holds for the primal problem, i.e. $V(\xi)+\int_0^T H_sdB_s\ge\xi$, ${\cal P}^\xi-$q.s. for some $H\in{\cal H}^\xi(\mathcal P)$.
\end{Theorem}

\textbf{Proof}
If we define $\widetilde E_n:=\left\{\omega\in\Omega \ : \ (\overline{\P}^n)^{\theta,\omega}=\nu(\omega)\right\}$ and then recursively
$$E_0:=\widetilde E_0\text{ and for all $n\geq 1$, }E_n:=E_n\bigcup\widetilde E_{n-1},$$
then $E_n$ is an increasing sequence such that $\mathbb P(E_n)\underset{n\rightarrow +\infty}{\longrightarrow} 1$. We can then use the $\mathcal F_\tau$-measurable kernel $\nu^n$ to define a probability measure $\widetilde P^n$ exactly as in the proof of Proposition \ref{pb}. We can then use exactly the same arguments as in our previous proofs.
\begin{flushright}$\Box$\end{flushright}

\end{document}